\documentclass[aps,twocolumn,prb,floatfix]{revtex4-2}
\usepackage{epsfig}
\usepackage{dcolumn}
\usepackage{bm}
\usepackage{amsmath}
\usepackage{tikz}
\usepackage{graphicx, graphics, color}
\usepackage{natbib}
\usepackage{mathtools}
\usepackage[colorlinks=true,citecolor=blue]{hyperref}

\ifpdf
	\DeclareGraphicsExtensions{.pdf,.png,.jpg,.mps}
\else
	\DeclareGraphicsExtensions{.eps}
\fi

\newcommand{\be}{\begin{equation}}
\newcommand{\ee}{\end{equation}}
\newcommand{\beq}{\begin{eqnarray}}
\newcommand{\eeq}{\end{eqnarray}}

\graphicspath{{./figs/}}

\begin{document}

\title{Puzzling superconductivity in strontium ruthenate: then and now}
\author{James F. Annett}
\affiliation{H. H. Wills Physics Laboratory, University of Bristol, Royal Fort, Tyndall Avenue, Bristol BS8 1TL, United Kingdom}


\author{Karol I.\ Wysoki\'{n}ski}
\affiliation{Institute of Physics, M.~Curie-Sk{\l}odowska University, pl.~M.~Curie-Sk{\l}odowskiej 1, 20-031 Lublin, Poland}

\date{\today}

\begin{abstract}

Strontium ruthenate is a very interesting low temperature superconductor with relatively simple normal state and very puzzling superconducting state properties. Despite over thirty years of intensive research, even the precise symmetry of the order parameter is not known. In this paper we briefly review
some of the key developments, focussing on some of the newer experiments, especially those which challenged the previous understanding that it is a spin triplet odd parity chiral p-wave superconductor.  Since Sr$_2$RuO$_4$ now appears most probably a
spin singlet superconductor we have calculated the Kerr effect for one candidate $d$-wave
pairing state. We find that for the chiral $d$-wave $E_g$ pairing state 
the calculated Hall conductance is non-zero and qualitatively similar
to that found earlier under the assumption of a chiral p-wave $E_u$ order parameter.
We characterize the Hall conductance in relation to two sum rules, one
related to Berry curvatures in the Bogoliubov quasiparticle states, and the 
other $f$-sum rule indicating the presence 
of inter-orbital pairing. 
\end{abstract}
\maketitle

\section{Introduction}

The discovery of superconductivity in cuprates, also known as high temperature superconductors, by Bednorz and Mueller in 1986~\cite{muller1986} led to
a major new ``superconducting paradigm''. Previously, according to the ``Matthais rules'', metallic oxides were not considered suitable candidate materials for superconductivity. Since the discovery of cuprate superconductivity, other major
classes of superconducting oxide materials have also been found, including cobaltates iridates and nickelates, as well as the closely related class of iron-based pnictides. 
Among these new oxide superconductors one of the most intriguing is 
the material Sr$_2$RuO$_4$, discovered to be a $1.5$K superconductor in 1994~\cite{maeno1994}.
Initially, there was the hope that comparison of this material to the cuprates might lead to a better understanding of the essential ingredients of the high-$T_c$ superconductivity in cuprates. However, there are key differences, and it turns out that the superconductivity in Sr$_2$RuO$_4$ has its own unique challenges and 
puzzles, which have taken many decades of research to understand.

The most obvious similarity between Sr$_2$RuO$_4$ and the cuprates is that
both systems possess the same crystallographic structure. Both Sr$_2$RuO$_4$
and the cuprate parent compound La$_2$CuO$_4$ possess the same $I4/mmm$
layered perovskite structure, of La$_2$NiO$_4$ type.
However, it soon became clear that both materials are very different in various respects. For example, the normal state of strontium ruthenate is a well-behaved anisotropic Fermi liquid in contrast to cuprates which have a marginal or non-Fermi liquid state above $T_c$. Furthermore, pure stoichiometric La$_2$CuO$_4$ is an antiferromagnetic insulator, and cuprate superconductivity only appears when the system is ``doped'', for example La$_{2-x}Ba_x$CuO$_4$  as discovered by Bednorz and Muller~\cite{muller1986}.
In contrast, for Sr$_2$RuO$_4$, the superconductivity
is present only in the stoichiometric pure material and doping 
is found to rapidly suppress superconductivity. The 'infinite layer' cubic perovskite SrRuO$_3$ and the bilayer Sr$_3$Ru$_2$O$_7$ are also non-superconducting. 
Therefore, unlike the cuprates, nickelates or cobaltites, there does not appear to be a wider class of superconducting ruthenate materials other than the unique case of pure  Sr$_2$RuO$_4$. 
However, the superconducting state of all the above classes of oxide systems as well as Sr$_2$RuO$_4$ seem to be unconventional, probably linked to magnetic fluctuations, rather than driven by classic BCS-type electron-phonon pairing. It is worth mentioning that the alloys  Sr$_2$Ir$_{1-x}$Ru$_x$O$_4$ were studied by R. J. Cava and coworkers~\cite{cava1994}. Interestingly,  for $ x=1$ the resulting compound is strontium ruthenate, while for very small $x$ (diluted limit) the Ru ions were found to act as magnetic impurities.

Evidence for unconventional superconductivity in Sr$_2$RuO$_4$ emerged at an early date, for example, in the strong sensitivity of $T_c$ to non-magnetic impurities~\cite{mackenzie2003}. A theoretical proposal by Rice and Sigrist, 
suggested that Sr$_2$RuO$_4$ could be a spin triplet superconductor analogous to
the ABM phase of superfluid He-3~\cite{rice1995}, with the order parameter $\Delta(k)\propto k_x\pm ik_y$. This time-reversal symmetry-breaking (TRSB) order parameter was consistent with muon spin relaxation experiments~\cite{Luke1998} and also with the Knight shift~\cite{ishida1998}. Further evidence for TRSB came with Kerr effect measurements by Xia {\it et al.}~\cite{xia2006}. 

 A major turning point was a more recent experimental measurement of the Knight shift performed in 2019~\cite{pustogow2019}, which, together with~\cite{chronister2021} challenged the spin-triplet character of the order parameter.  It was subsequently 
 acknowledged that the original Knight shift measurements~\cite{ishida1998} were affected
by sample heating~\cite{ishida2020}, and that when this was corrected 
for the results became
consistent with a spin singlet (eg $s$-wave or $d$-wave) pairing state and inconsistent
with the original spin triplet proposal of Rice and Sigrist~\cite{rice1995}.
Similar conclusions were reached independently using neutron scattering~\cite{Petsch2020}.

Before we discuss the experiments
relating to the superconducting state of strontium ruthenate, we wish to make a few comments on the possible pairing mechanism in Sr$_2$RuO$_4$ in contrast to that in cuprate high temperature superconductors. The most general statement would be that the pairing mechanisms are not certain in either of these types of superconductors. A more detailed remark would be that there is a
significant difference in the normal state of the two classes of materials. In the cuprates the pairing emerges out of an anti-ferromagnetic Mott insulating state, while in Sr$_2$RuO$_4$ pairs evolve out of a correlated metal. The prevailing view is that in both systems the magnetic~\cite{spalek2022} and charge fluctuations~\cite{gingras2022} play the most important role, while phonons seemingly play a less relevant role, albeit one has to remember that cuprate superconductors show many different isotope effects~\cite{bussman2022}, which by definition require the couplings of the relevant degrees of freedom to phonons. For cuprates it is generally believed that the dominant interaction is a strong on-site Hubbard repulsion, $U$, with  superconductivity arising around a quantum critical point where the antiferromagnetic Mott insulating phase vanishes
\cite{Anderson1987,Maier2005}. In contrast, for Sr$_2$RuO$_4$ it is now understood that the normal state can be characterized as a ``Hund metal'' with both intra and inter-orbital Hubbard $U$ and inter-orbital exchange $J$ the relevant interactions\cite{Deng2016,Kaser2022}. This correlated 
Hund metal can become superconducting in several 
possible pairing states, depending on precise details of Fermi surface and interaction parameters\cite{Kaser2022,Romer2022}.

In this paper, we shall first review the main facts about the normal and superconducting state of strontium ruthenate in Section~\ref{sec:overview}. Due to the important role of the Knight shift measurements in the history of more than thirty years of struggle to identify the spin structure of the superconducting order parameter in Section~\ref{sec:knight-shift}, we discuss in some detail both experiments measuring the Knight shift: the first one from 1998 and the second from 2019. Our own approach to the Kerr effect is presented in Section~\ref{sec:trs-kerr}, where we calculate the frequency-dependent Hall conductivity $\sigma_{xy}(\omega)$ in time-reversal symmetry (TRS) breaking superconductors. 
In Section \ref{sec:res-kerr} we present new calculations of the Kerr
effect for a $d$-wave pairing state with TRSB.  Previous Kerr effect 
calculations, by ourselves and by others, had assumed a spin triplet state.
Here we present the calculated Hall conductivities $\sigma_{xy}(\omega)$ for the chiral spin-singlet state, as well as showing two  Hall effect sum rules,
one related to Berry curvatures of the Bogoliubov de Gennes quasiparticle bands and the other linked to the current-current
commutator $\langle [ \hat{j}_x,\hat{j}_y] \rangle$ evaluated in the TRSB
superconducting state. We finish with a summary and discussion of open questions in Section ~\ref{sec:summary}.

\section{Overview of Sr$_2$RuO$_4$ and brief summary of  early results} \label{sec:overview}

There are already many excellent reviews of the superconductivity in Sr$_2$RuO$_4$ in the literature~\cite{maeno2001,mackenzie2003,bergemann2003,kallin2012,maeno2012,kallin2016,mackenzie2017,huang2021,leggett2021,maeno2024}. Despite that, for the sake of clarity, here we recall some of the experimental facts which are important for the later
discussion in this paper.

First of all, this material is stoichiometric and extremely clean, with a very long transport mean-free-path. The superconductivity, however, is very fragile and only a very small amount of impurities, in the ppm range, 
is sufficient to totally suppress superconductivity~\cite{mackenzie1998}. 
For conventional $s$-wave BCS superconductors non-magnetic impurities have little effect on $T_c$ as a result of Anderson's theorem. Therefore, the rapid suppression of $T_c$ by non-magnetic impurities was already a clear signature
of unconventional Cooper pairing, either of triplet $p$ or singlet $d$ type. 
At that time, the band structure of the material was experimentally established~\cite{mackenzie1996} and agreed with DFT calculations. There were experimental hints about the importance of correlations, but within a well-defined Fermi-liquid normal state. 

 The proposal of Rice and Sigrist~\cite{rice1995} that strontium ruthenate could be a spin triplet superconductor was made very soon after its original discovery. 
They proposed that the pairing state of the Sr$_2$RuO$_4$ would be a solid-state analogue of the ABM phase of superfluid $^3$He~\cite{rice1995}. The motivation for this proposal relied on the few experimental facts known at that time. They noted that a related material SrRuO$_3$ was a ferromagnet, and that in the diluted limit, Ru$^{4+}$ in Sr$_2$Ir$_{1-x}$Ru$_X$O$_4$ alloy acts as a local $S = 1$ moment impurity in Sr$_2$IrO$_4$. 
A Fermi liquid with spin triplet fluctuations and tendency to ferromagnetism
suggested, in analogy to superfluid $3$-He, spin-triplet pairing. 
Rice and Sigrist~\cite{rice1995} also hinted that the best way to uniquely identify the spin structure is a measurements of Knight shift, which shows qualitatively different temperature dependence for singlet and triplet states. Indeed, the  Knight shift measured soon afterwards~\cite{ishida1998} was found to follow the dependence expected for spin triplet superconductors.  

The proposed spin-triplet pairing state became the standard picture of Sr$_2$RuO$_4$
for the next two decades.  Several independent experiments provided 
evidence that appeared to support this picture, notably the evidence for time reversal 
symmetry breaking (TRSB) found in muSR~\cite{Luke1998} and Kerr effect~\cite{xia2006} in the superconducting state.  On the other hand, some of the expected features
of the proposed chiral $p$-wave state were not observed, for example, the predicted
spontaneous edge currents were absent~\cite{Curran2014,Curran2023}.
Many other experimental results appeared to be contradictory~\cite{kivelson2020}, which makes unique identification of the superconducting state very difficult. For example, there are thermodynamic indications of the gap nodes, presumably vertical, $i.e.$ along the z-axis~\cite{hasinger2017}, but it is not clear where on the Fermi surface they are located. 

In the next section we review in more detail two key
experimental results which are relevant to
our understanding of the superconducting state in Sr$_2$RuO$_4$, focussing
specifically on 
the Knight shift~\ref{sec:knight-shift} and evidence for TRSB 
from the Kerr effect~\ref{sec:trs-kerr}.

\section{Review of Recent Results}
\subsection{ Knight shift} \label{sec:knight-shift}

The Knight shift measures the spin susceptibility in the superconducting state. It provides the information necessary for determining the spin state of the superconducting pairs. Early measurements of Knight shift~\cite{ishida1998} played a pivotal role in establishing the spin-triplet structure of Cooper pairs in strontium ruthenate. To understand the relation between the structure of Cooper pairs and spin susceptibility as measured by the Knight shift, it is important to recall that for spin triplet superconductors, spin susceptibility is a tensor with entries dependent on the relative orientation of
the spin of the triplet Cooper pair and the magnetic field~\cite{leggett1975}. In particular, the spin susceptibility has the following tensor form
\begin{equation}
  \hat{\chi}_s(T) =  \chi_n \left( \begin{array}{ccc} 1 & 0 & 0 \\
  0 & 1 & 0 \\
  0 & 0 & Y(T)  \end{array}\right) \label{eq:chi-a}
\end{equation}
for the superconducting state with the $d$-vector triplet gap 
function of the type  
\begin{equation}
 {\bf d}({\bf k}) = (\sin{k_x} \pm i\sin{k_y})\hat{\bf e}_z=
 \Delta_{\bf k}\hat{\bf e}_z, 
 \label{eq:chiral-a}
\end{equation}
corresponding to $L_z=\pm 1 $, $S_z=0$ spin triplet Cooper pairs.
Above, we have denoted by $\chi_n$ the normal state Pauli spin susceptibility and by $Y(T)$ the Yoshida function. This function is temperature-dependent and given by
\begin{equation}
    Y(T)=\frac{1}{2k_BT} \int_0^\infty d\epsilon_{\bf k} \rm{sech}^2
    \left(\frac{E_{\bf k}}{2k_BT} \right), 
\end{equation}
where $T$ is temperature and $E_{\bf k}=\sqrt{\epsilon_{\bf k}^2+|\Delta_{\bf k}|^2}$. The Yoshida function equals $1$ for $T=T_c$ and drops towards zero at $T\rightarrow{0}$. Thus, the spin susceptibility measured as a function of temperature remains constant with decreasing temperature from the normal state down to zero for a magnetic field in $ab$ plane and decreases below $T_c$ for a magnetic field along the $c$-axis. 

The early measurements of Knight shift in an $ab$-magnetic field~\cite{ishida1998} reported behaviour consistent with the state (\ref{eq:chiral-a}), namely
a constant $ab$-plane spin-susceptibility in both the normal and superconducting states. It is important to recall some parameters of measurements, as their values will be important in the following. The authors measured $^{17}$O Knight shift for all inequivalent oxygen sites. In the temperature range 1.4 - 200 K, the Knight shift was measured at 58.1MHz (with an external magnetic field $H_{ext} < 100 kOe$), while in the range $15mK - 1.8 K$ the 
measurements were carried out at lower frequencies and fields; namely  3.8MHz and $H_{ext} < 6:5 kOe$. The duration of the pulses was constant and of standard length.

Later on, Murakawa {\it et al.} ~\cite{murakawa2004} measured the $^{101}$Ru-Knight shift of  Sr$_2$RuO$_4$ in a superconducting state under the influence of a magnetic field parallel to the c-axis.   In contradiction to expectations from Eqs. (\ref{eq:chiral-a}) and (\ref{eq:chi-a})  they found that its value is also unchanged from the normal state one below $T_c$. The technique used was nuclear-quadrupole-resonance under magnetic fields parallel to the RuO2 plane (ab direction). The magnetic field was 550 Oe or greater, and the lowest temperature was 97mK. This finding was consistent with other possible triplet states predicted for strontium ruthenate~\cite{annett1990}, namely
\begin{eqnarray}
  {\bf d}({\bf k}) &=& (\sin{k_x},\sin{k_y},0)   \label{eq:nonchiral} \\
   {\bf d}({\bf k}) &=& (\sin{k_y},-\sin{k_x},0),  \nonumber
\end{eqnarray} 
for which the expected spin susceptibility tensor reads~\cite{leggett1975,annett1990}
\begin{equation}
  \hat{\chi}_s(T) =  \frac{1}{2} \chi_n \left( \begin{array}{ccc} 1+Y(T) & 0 & 0 \\
  0 & 1+Y(T) & 0 \\
  0 & 0 & 2  \end{array}\right). \label{eq:chi-bc}
\end{equation}
Thus one NMR experiment indicated the state with $d$-vector along the $c$-axis and another
with it in $ab$-plane. In view of the fact that application of the magnetic field may rotate the $d$-vector, one plausible solution to the puzzle would be the rotation of the $d$-vector by a small applied field~\cite{annett2008}. Such an effect is known from the studies of spin-triplet superfluid $^3$He-A ~\cite{leggett1975}. 
However, in contrast to the isotropic fluid $3$-He, in a solid material the orientation of the $d$-vector order parameter is determined by spin-orbit coupling. 
Therefore, whether or not the $d$-vector would rotate in the small $c$-axis magnetic
fields of the experiment is dependent on the precise parameter values assumed~\cite{annett2008}.

This picture changed dramatically with
more recent measurements~\cite{pustogow2019,ishida2020} of the Knight shift. These have revealed some limitations of the previous measurement technique and have shown that the spin susceptibility does not drop down at $T_c$. The crucial point is the length of the measurement pulse and the power deposited in the sample. The new experiments~\cite{pustogow2019} measuring the $^{17}$O Knight shift carefully monitored the power deposited during the measurements. It turned out that for long pulses and large power deposited in the sample, the results agreed with earlier ones~\cite{ishida1998} and spin susceptibility at small fields remained at its normal state value. However, at a smaller power, the spin susceptibility dropped below its normal state value at the transition temperature. Careful analysis of experimental aspects has been performed by K. Ishida {\it et al.} \cite{ishida2020}. They confirmed the results of Pustogow {\it et al.}~\cite{pustogow2019} and commented on their original results: ``We conclude that the previous results of the invariance of the Knight shift in the SC state were due to instantaneous destruction of superconductivity by the RF pulses.''

These results were confirmed by an
entirely independent measurement of the spin-susceptibility from 
neutron scattering. In 2020 Petsch {\it et al.}~\cite{Petsch2020}
found a strong decrease in spin susceptibility 
below $T_c$, significantly improving on earlier measurements
which had a larger error bar~\cite{Duffy2000}.
Essentially, the same conclusions have been obtained from the precise $\mu$SR measurements~\cite{matsuki2026}, sometimes referred to as muon Knight shift measurements. 

Taken at face value, these new spin susceptibility measurements appear to rule
out the possibility of a spin triplet pairing state.  But the situation
is complicated by spin-orbit coupling, which means that the 
simple forms of the spin susceptibility given in Eqs.~\ref{eq:chi-a}
and ~\ref{eq:chi-bc} are not precisely correct. Nevertheless, Gupta {\it et al}~\cite{Gupta2020} 
calculated the spin-susceptibility assuming a spin triplet pairing state
of the form of Eq.~\ref{eq:nonchiral} and found that the suppression
of susceptibility below $T_c$ was not consistent with the experiments for reasonably estimated values of the spin-orbit interaction parameter. Better agreement
was found under the assumption of $d$ wave pairing, as found by Romer {\it et al.}~\cite{romer2019} and by Gupta {\it et al.}~\cite{Gupta2022}.  
However, unfortunately, the observed spin-susceptibility is consistent with a wide range of candidate $d$-wave pairing states, and so while these measurements
now clearly rule out spin-triplet pairing states, they do not narrow down the choice of order parameter to a unique state. If we take the requirement of a pairing state with TRSB as also experimentally required, then several possible pairing states remains under consideration, including 
$d_{xz}+id_{yz}$, $d+is$ and $d+ig$~\cite{kivelson2020}.

\subsection{TRSB, muSR and Kerr effect}\label{sec:trs-kerr}

Time reversal symmetry breaking (TRSB) has been reported in a number of
superconductors~\cite{ghosh2021,wysokinski2019}. For Sr$_2$RuO$_4$, TRSB was first reported by Luke {\it et al.}~\cite{luke1998}, observed as a change in muon spin relaxation rates as the material becomes superconducting. Subsequently, a polar Kerr effect, a rotation of the plane of linear polarized light reflected from the surface~\cite{xia2006} was also found, with approximately a $65$ nano-radian rotation of the plane of the optical polarization. In both cases, the effect occurred at the same critical temperature, $T_c$ as superconductivity, implying that the TRSB observed was an intrinsic property of the superconducting state, rather than a separate magnetic phase or a result of magnetic impurities or inclusions. The sign of the Kerr effect signal could be ``trained'' by sweeping the field up or down in field-cooled samples. 
The Kerr signal also implies that if there are domains of opposite sign of TRSB within the superconducting phase, these must be at least the size of the optical spot size of the interferometer laser, $\approx 25 \mu$m. Both Luke {\it et al.}~\cite{luke1998}
and Xia {\it et al.}~\cite{xia2006} noted that their observations would be consistent with the chiral p-wave state expected at that time.

A general theory of the polar Kerr effect in clean unconventional 
superconductors was originally developed by Capelle, Gross and Gyorffy~\cite{capelle1997,capelle1998}.   
Taylor and Kallin~\cite{taylor2012} calculated the Kerr
effect in a simplified 2D two-band model of chiral p-wvave Sr$_2$RuO$_4$.
In this approach, the effect arises from interband transitions involving time-reversal symmetry-breaking chiral Cooper pairs. The crucial role of inter-orbital coupling providing a non-zero value of the Kerr effect has been noted independently in Ref.~\cite {wysokinski2012} using the three-dimensional multi-orbital model of strontium ruthenate and calculating the difference in the power absorbed by left and right circularly polarised light~\cite{capelle1998}.
The calculations were later extended by  Gradhand {\it et al.}~\cite{gradhand2013} (correcting
an error in~\cite{wysokinski2012}). 

In both the Taylor and Kallin and Gradhand {\it et al.} approaches, the Kerr effect derives from a finite
frequency Hall conductance, which can be written in the form
\begin{eqnarray}
    {\rm{Im} }[ \sigma_{xy}(\omega) ]=   \frac{e^2}{\hbar} 
    \frac{\pi}{2 \hbar\omega V} &&
     \sum_{{\bf k},n,n^\prime} {\rm Im} ( M_{xy}-M_{yx}) \times \nonumber \\
    &&  [ 1- f(E^{n^\prime}({\bf k}))]  f(E^{n}({\bf k})) 
    \times \nonumber \\
    && \delta(E^{n^\prime}({\bf k})-E^{n}({\bf k})-\hbar \omega)
    \label{eq:imhall}
\end{eqnarray}   
where $E^{n^\prime} ({\bf k})$ and $E^{n} ({\bf k})$ are 
quasiparticle energies of the
Bogoliubov de Gennes equations
\begin{equation}
  \left(  \begin{array}{cc} H_{\bf k}  & \Delta_{\bf k} \\
    \Delta^*_{\bf k} &  -H^*_{\bf k}  \end{array} \right)      
  \left( \begin{array}{c}  u^n_{\bf k}   \\
    v^n_{\bf k}  \end{array}  \right)  =
    E^n({\bf k})  \left( \begin{array}{cc} u^n_{\bf k}   \\
    v^n_{\bf k}   \end{array}  \right)     
\end{equation}
and $f^{n\prime} ({\bf k}) = f(E^n({\bf k}))$ is the Fermi function.
The matrix elements are of the form
\begin{equation}
  M_{xy} =    \langle  n {\bf k} \mid \hat{j}_x \mid  n^\prime {\bf k} 
     \rangle \langle  n^\prime {\bf k} \mid \hat{j}_y
     \mid  n {\bf k}  \rangle  
\end{equation}
and similarly for $M_{yx}$, and where the current operator is
\begin{equation}
  \hat{\bf j} =\left( \begin{array}{cc}  \nabla_{\bf k} H_{\bf k}  & 0 \\
    0 &  \nabla_{\bf k} H_{\bf k}  \end{array}  \right) .      
\end{equation}
In Sr$_2$RuO$_4$ the quantities $H({\bf k})$ and $\Delta({\bf k})$
are generally $3\times 3$  matrices, corresponding to the three 
bands at the Fermi surface. If spin-orbit coupling is also included in the model, they become $6\times 6$  matrices.
The real part of the Hall conductance,  
${\rm{Re} } [\sigma_{xy}(\omega)]$, is obtained from the imaginary part, as usual, by the Kramers-Kronig relations. 
The inter-band contributions  arise in this formalism from
the sum over the initial and final states, $n$ and $n^\prime$, 
and the chirality arises from the inter-band 
scattering matrix elements $M_{xy}$. Perhaps surprisingly, the 
dominant frequencies, $\omega$, present in the Hall spectrum are not 
those of the pairing gap $\Delta$, of order $10^{-4}$eV for a 1.5K
superconductor such as Sr$_2$RuO$_4$, but are at much higher energies related to the band
energy differences, which for Sr$_2$RuO$_4$ are around $0.05-0.2$eV. Note that the actual laser frequency used in the experiments of Xia {\it et al.}~\cite{xia2006} is at an even larger optical energy of about $0.8$eV, and so some extrapolation is needed to estimate the actual observed Hall angle
from the calculated spectrum. Within the uncertainties of this
extrapolation it is estimated that, for the assumed chiral p-wave pairing state, the results were comparable to the $65$nrad rotation angle observed~\cite{gradhand2013}. 

 Important differences between chiral inter-band and intra-band 
 pairing states were discussed by Mineev~\cite{mineev2007,mineev2012,mineev2014}.
 In particular, he noted that an intrinsic Kerr effect cannot exist
 in a single band system.  Further confirmation of this point was
 obtained by relating the Hall effect to Berry curvatures of the Bogoliubov quasiparticle Bloch states $(u^n_{\bf k},v^n_{\bf k}) $ 
 via the sum rule
 \begin{equation}
     \int_0^\infty \frac{ {\rm{Im} } \sigma_{xy}(\omega)}{\omega}
     d \omega = - \frac{e^2\pi}{2\hbar} \sum_{n,\bf k} 
      \Omega^z_n({\bf k} ) f(E^n({\bf k})) .
      \label{eq:berry}
 \end{equation}
 The Berry curvatures, $\Omega^z_n({\bf k})$ (defined in Ref.~\cite{gradhand2014})
 are automatically zero in a single band, hence explaining the absence of a Kerr signal in one-band models. 
 Using the same three-band model of Sr$_2$RuO$_4$ as used previously~\cite{wysokinski2012,gradhand2013} it was found that the sum rule
 is obeyed fully for the model chiral p-wave pairing state~\cite{gradhand2014}.

Note that the above calculations were carried out for the case
of an ideal periodic crystal. Impurity scattering 
in a chiral p-wave superconductor provides an alternative 
possible source of a non-zero Kerr effect~\cite{goryo2007,lutchyn2009,konig2017}.
In this case a Kerr effect should be present even in a single-band superconductor.

\section{Kerr effect for a model chiral d state}\label{sec:res-kerr}


As discussed earlier, the new experiments on the Knight shift are 
incompatible with the previously assumed chiral p-wave spin-triplet state. 
Thus, it is important to check if the Kerr effect calculated for the chiral state of a singlet character will be able to describe the existing results on the Kerr effect~\cite{xia2006}. In the literature, there are several 
different proposals~\cite{romer2019,kivelson2020} for the singlet 
states expected to be compatible with recent, correct measurements 
of the Knight shift. Here we shall present the results of our 
calculations using a three-dimensional model of the strontium ruthenate 
spectrum and chiral $d_{xz}+id_{yz}$ symmetry.  This is the only 
$d$-wave pairing state with a symmetry required
two-fold degeneracy at $T_c$, necessary for TRSB without two
separate phase transitions, and is the 
closest analogue to the previously studied chiral $p$-wave,
$p_x+ip_y$ state for which the Kerr effect
was found earlier.


Our model chiral $E_g$ d-wave pairing state of Sr$_2$RuO$_4$ 
is of the form examined previously by Gupta {\it et al.}~\cite{Gupta2022}.
The gap function is assumed to be of the form
\begin{eqnarray}
    \Delta_{mm^\prime}({\bf k}) & = & \sin{(\frac{k_z c}{2})} \left(  \Delta_{mm^\prime}^{x} \sin{(\frac{k_x a}{2})}\cos{(\frac{k_y a}{2})} \right. \nonumber \\
     & +& \left.  \Delta_{mm^\prime}^{y} \cos{(\frac{k_x a}{2})}\sin{(\frac{k_y a}{2})}  \right)
     \label{BdeG}
\end{eqnarray}
where the ${\bf k}$-dependent basis functions correspond to the simplest
allowed functions transforming according to the $E_g$ irreducible
representation of the $D_{4h}$ point symmetry group in a body centred tetragonal $I4/mmm$ space group.    
Note that the symmetry required line node in the plane $k_z=0$ implies that
generally out of plane interactions are required to stabilise 
such a pairing state, and therefore this pairing state is not possible in
purely two-dimensional models of a single superconducting
Sr$_2$RuO$_4$ plane. (Excluding more complex proposals involving 
inter-orbital pairing combined with 
strongly ${\bf k}$ dependent spin-orbit interactions~\cite{Suh2020}). 

\begin{figure}[t]
\centerline{\includegraphics[width=0.85\linewidth]{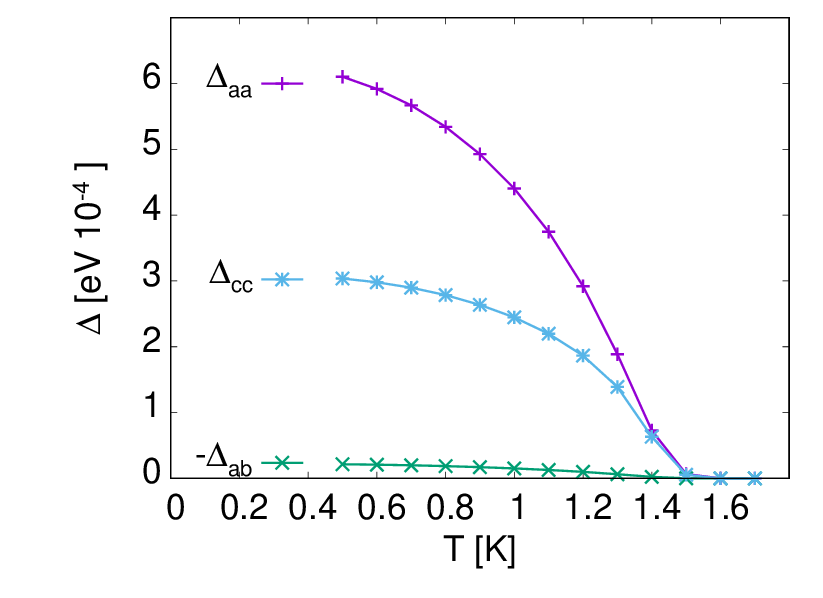}}
\caption{Temperature-dependent gap parameters in the model
$E_g$ pairing state considered here. We denote the three Ru $d$-orbitals as 
$a$ ($d_{yz}$), $b$ ($d_{zx}$) and $c$ ($d_{xy}$) respectively. 
(color online)  }
\label{fig:rys3}  
\end{figure} 

To develop a simple model $E_g$ $d$-wave pairing state we assume an effective pairing interaction of the form
\begin{equation}
    U_{mm^\prime} = \left( \begin{array}{ccc} 
     U_{aa} & U_{ab} & 0 \\
     U_{ab} & U_{bb} & 0 \\
     0    & 0 & U_{cc}
    \end{array} \right), 
\end{equation}
where here we are denoting the three $Ru$ d orbitals as $a=d_{yz}$, $b=d_{zx}$, $c=d_{xy}$.
In real-space this assumed pairing interaction is chosen to be along the nearest body-centred
neighbour directions ${\bf R}=(\pm a/2,\pm a/2, \pm c/2)$, 
which results in the $E_g$ gap function basis functions corresponding to Eq.~\ref{BdeG}.
Using the same three-dimensional tight-binding parameters as Gradhand {\it et al.}~\cite{gradhand2013} and Gupta {\it et al.}~\cite{Gupta2022}, but omitting spin-orbit coupling, we solve the self-consistent Bogoliubov de Gennes equations.
Fig.~\ref{fig:rys3}  shows the results for the specific choice 
$U_{aa}=U_{bb}= 1.60 t$, $U_{ab}= 0.60 t$, $U_{cc}= 0.3107t$, where $t=0.08162$ eV is the nearest neighbour hopping parameter in the dominant $\gamma$-band.
The gap parameters not shown in Fig.~\ref{fig:rys3} are either
zero or equivalent in symmetry to the ones shown. For example,
\begin{eqnarray}
     \Delta_{cc^\prime}^{y}  &=& i \Delta_{cc^\prime}^{x}  \nonumber \\
     \Delta_{bb^\prime}^{y}  &=& i \Delta_{aa^\prime}^{x}.
\end{eqnarray}
We find that these symmetries emerge directly from
the numerically self-consistent solution of the Bogoliubov de Gennes equations. They confirm that the solution found is in the  $(1,i)$ chiral symmetry $E_g$ state, since the gap parameters are found to be invariant
under the combined $C_4$ rotation about the $z$-axis and a $\pi/2$ gauge
transformation~\cite{annett1990}. 

Note that the choice of orbital-dependent parameters is not unique, and it is possible
to vary the individual parameters considerably while retaining a constant
value of $T_c=1.5$K. In Fig.~\ref{fig:rys3}, the value of $U_{aa}$ was increased
compared to the value used by Gupta {\it et al.}~\cite{Gupta2022} in order
to make the gap larger on the $\alpha$ and $\beta$ bands of the Fermi
surface. This was done because the calculated Hall effect is found to
be dominated almost exclusively by the $\alpha$ and $\beta$ band contributions, with a negligible contribution from the $\gamma$-band. 

\begin{figure}[t]
\centerline{\includegraphics[width=0.85\linewidth]{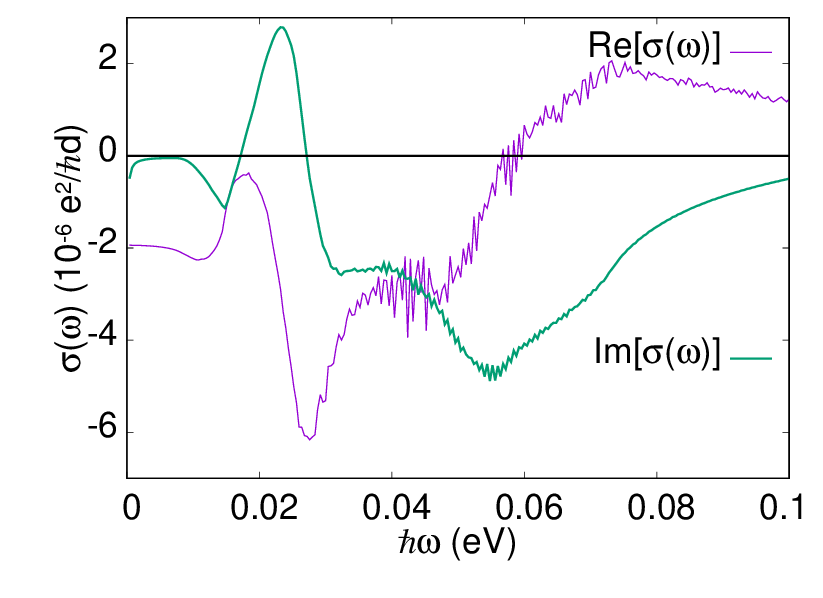}}
\centerline{\includegraphics[width=0.85\linewidth]{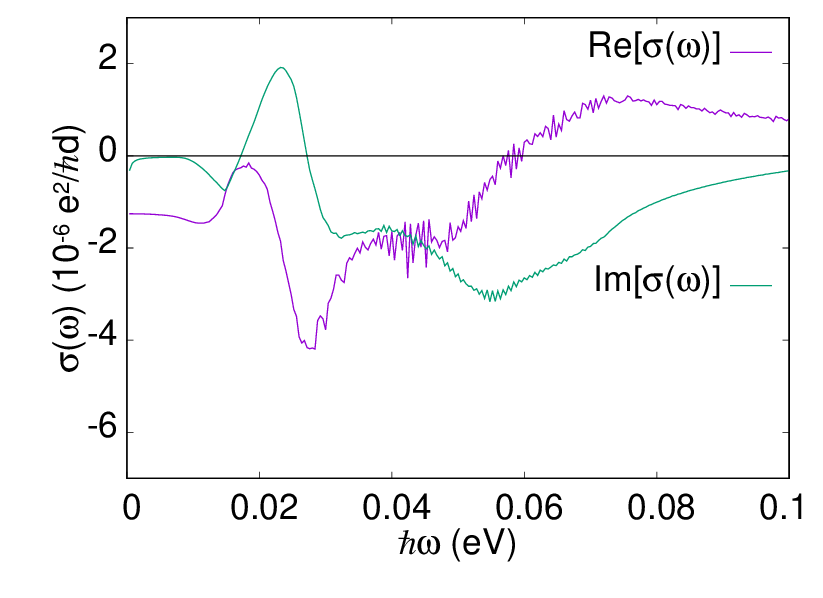}}
\caption{Calculated Hall conductance $\sigma_{xy}(\omega)$ for the 
chiral $E_g$ ($d_{xz}+i d_{yz}$) pairing state. 
The upper panel is $T=0.5$K, and the lower panel 
for $T=1$K, showing the reduction of the magnitude of the 
Hall effect spectrum as the
pairing gap reduces to zero as $T \rightarrow T_c$. Note the
scale is $10^{-6} e^2/\hbar d$, where $d=c/2$ is the interlayer
separation in Sr$_2$RuO$_4$. (color online)  }
\label{fig:rys4}  
\end{figure} 

The calculated Hall conductance is shown in Fig.~\ref{fig:rys4} for temperatures $T=0.5$K and $1$K. Clearly, there is a similar spectrum in both cases, but the overall magnitude is smaller at the higher temperature of $T=1$K. 
The real part $ {\rm Re}[ \sigma_{xy}(\omega)] $  is 
derived from the imaginary part
by the Kramers-Kronig relation 
\begin{equation}
    {\rm Re}[ \sigma_{xy}(\omega)]  = \frac{2}{\pi} {\cal P} \int_0^\infty \frac{\omega^\prime }{\omega^{\prime 2}-\omega^2} {\rm Im}[\sigma_{xy}(\omega')] d\omega^\prime ,
    \label{eq:kramers}
\end{equation}
where $\cal{P}$ denotes the principal part integral. This gives
\begin{eqnarray}
    {\rm{Re} }[ \sigma_{xy}(\omega) ]&=&   \frac{e^2}{\hbar} 
    \frac{1}{V} 
     \sum_{{\bf k},n,n^\prime} {\rm Im} ( M_{xy}-M_{yx}) \times \nonumber \\
    &&  \frac{[ 1- f(E^{n^\prime}({\bf k}))]  f(E^{n}({\bf k})) }
   {(E^{n^\prime}({\bf k})-E^{n}({\bf k}))^2-(\hbar \omega)^2}
    \label{eq:rehall}
\end{eqnarray} 
and where the imaginary part of the Hall conductance is obtained from Eq.~\ref{eq:imhall}~\cite{gradhand2013}.

In the calculated spectrum shown in Fig.~\ref{fig:rys4} the real part 
${\rm{Re} }[ \sigma_{xy}(\omega) ]$ 
has a finite value at zero frequency, $\omega=0$. From the Kramer-Kronig relation given above this value is
\begin{equation}
    \int_0^\infty \frac{Im[\sigma_{xy}(\omega)]}{\omega} d\omega
     = \frac{\pi}{2} Re[\sigma_{xy}(0)] .
     \label{eq:sumrule0}
\end{equation}
The integral is the same one as the integral giving the sum rule related to the
Berry curvature sum, Eq.~\ref{eq:berry} and so the 
value of  $Re[\sigma_{xy}(0)]$ is a direct measure of the 
Berry curvatures in the Bogoliubov quasiparticle bands in the 
superconducting state~\cite{gradhand2014}. 
However, it should be noted that the form of the Hall conductance
given in Eqs.~\ref{eq:imhall} and~\ref{eq:rehall} is not valid in the
strictly zero frequency, or $d.c.$, limit, because screening supercurrent flows
are omitted from the calculation of the optical response function. 

\begin{figure}[t]
\centerline{\includegraphics[width=0.85\linewidth]{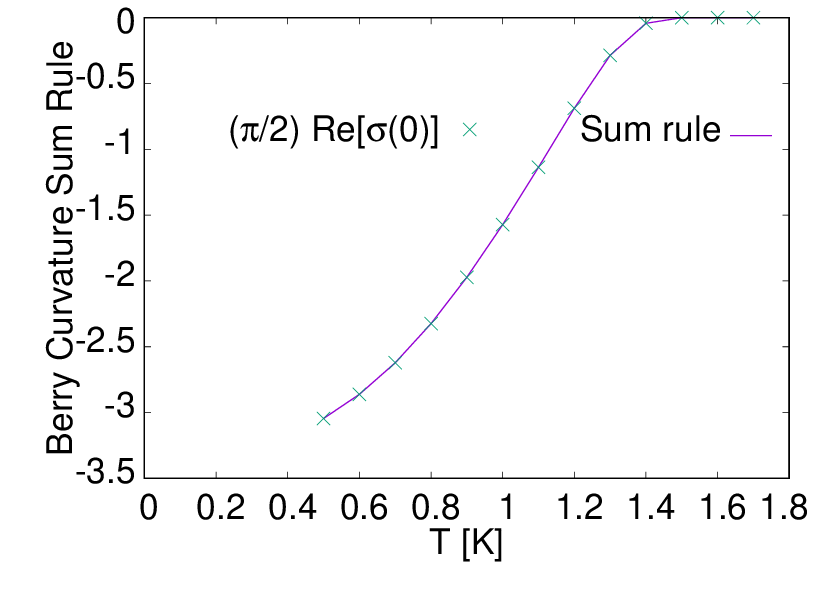}}
\centerline{\includegraphics[width=0.85\linewidth]{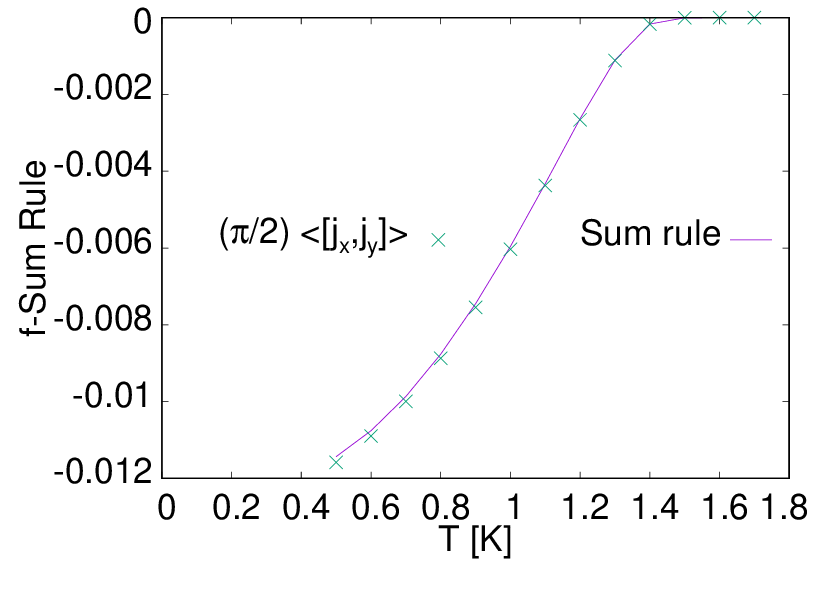}}
\caption{Top panel: Berry curvature sum rule compared to zero
frequency conductance $Re[\sigma_{xy}(0)]$. Lower panel
f-sum rule compared to the calculated current commutator, 
$\langle [\hat{j}_x,\hat{j}_y] \rangle$.
Here, for simplicity, we use natural units $e=\hbar=1$. (color online)}
\label{fig:rys5}  
\end{figure} 

It is illuminating to examine the change in the Hall spectrum as a function
of temperature.   In Fig.~\ref{fig:rys4} we can see that the overall 
magnitude of the spectrum decreases at $T \rightarrow T_c$, while the
range of frequencies for which ${\rm Im}\sigma_{xy}[(\omega)]$ is non-zero
remains approximately constant.  We can also see that at
$T=0.5$K   the low frequency limit ${\rm Re}[\sigma_{xy}(0)]$ is approximately
$2  \times 10^{-6} e^2/\hbar d$ (where $d=c/2$ is the inter-layer
separation in Sr$_2$RuO$_4$). This value has become about 
$1 \times 10^{-6} e^2/\hbar d$ at $T=1$K.  The upper
panel of Fig.~\ref{fig:rys5} shows the full temperature dependence
of this value, showing a smooth decrease towards zero at $T=T_c$. 
The numerical evaluation of the frequency sum rule integral over the calculated 
spectrum is found to agree very well with the values of
${\rm Re}[\sigma_{xy}(0)]$, confirming the numerical stability of the results shown.

In the lower panel in Fig.~\ref{fig:rys5} we show a second sum rule, 
which is related to the $f$-sum rule~\cite{Lange1999,Zhang2026}
\begin{equation}
    \int_0^\infty \omega Im[\sigma_{xy}(\omega)] d\omega
     = \frac{e^2}{\hbar^3 V}  \frac{\pi}{2} \langle [ \hat{j}_x,\hat{j}_y ] \rangle .
     \label{eq:fsum}
\end{equation}
The proof of this relation in the superconducting state is straightforward
starting from Eq.~\ref{eq:imhall}. First, we note that inside the integrand
of Eq.~\ref{eq:imhall} the product of Fermi functions $[ 1- f(E^{n^\prime}({\bf k}))]  f(E^{n}({\bf k})) $ can be replaced by just $f(E^{n}({\bf k}))$,
because the product term,
$f(E^{n^\prime}({\bf k})) f(E^{n}({\bf k})) $, is symmetric in the states
$n$ and $n^\prime$ and therefore sums to zero~\cite{gradhand2014}.
Carrying out the frequency integral in Eq.~\ref{eq:fsum} we are left with terms like
\begin{eqnarray}
 \sum_{n^\prime} M_{xy}   &=&   \sum_{n^\prime}  \langle  n {\bf k} \mid \hat{j}_x \mid  n^\prime {\bf k} 
     \rangle \langle  n^\prime {\bf k} \mid \hat{j}_y
     \mid  n {\bf k}  \rangle    \nonumber \\
     &=&   \langle  n {\bf k} \mid \hat{j}_x  \hat{j}_y
     \mid  n {\bf k}  \rangle  ,
\end{eqnarray}
where we are using completeness of the states $\mid n^\prime {\bf k}\rangle$.
Hence, the sum rule yields
\begin{eqnarray}
    \int_0^\infty \omega Im[\sigma_{xy}(\omega)] d\omega
    & = & \frac{e^2}{\hbar} 
    \frac{\pi}{2 \hbar^2 V}  \sum_{n {\bf k} }
     \langle  n {\bf k} \mid [\hat{j}_x,  \hat{j}_y]  \mid  n {\bf k}  \rangle  \nonumber \\
   && \times      f(E^{n}({\bf k})) ,
\end{eqnarray}
where the states $\mid  n {\bf k}  \rangle$ are the eigenvectors of the
Bogoliubov Hamiltonian. 
The confirmation of this relation numerically in our calculated Hall spectra
is shown in the lower panel of Fig.~\ref{fig:rys5}.  Note that for the specific pairing state considered here, the numerical magnitude of this
sum is much smaller than found for the Berry curvature sum rule. 
That is because this sum rule is in effect a measure of the presence of the inter-orbital pairing $\Delta_{ab}$ in our gap model.  The small value of
the sum rule integral shown in Fig.~\ref{fig:rys5} is a consequence of the small value of $\Delta_{ab}$ shown in Fig.~\ref{fig:rys3}.
If we choose a value of $U_{ab}=0$ this forces $\Delta_{ab}=0$ and in this case the commutator
$\langle [ \hat{j}_x,\hat{j}_y ] \rangle$ vanishes exactly. 

Finally, we note that the experimental
measurements of the Kerr angle are made at photon energies of $0.8$eV~\cite{xia2006}, considerably larger than the frequency range shown in Fig.~\ref{fig:rys4}. 
Using the Kramers-Kronig relation, we can see that for large frequencies
\begin{eqnarray}
     Re[\sigma(\omega)] & \sim &\frac{1}{\omega^2}  
      \int_0^\infty  \omega^\prime Im[\sigma_{xy}(\omega^\prime)] d\omega^\prime . \\
       & \propto&   \frac{1}{\omega^2}   \langle [ \hat{j}_x,\hat{j}_y] \rangle . \nonumber
\end{eqnarray}
Therefore we can observe that the Berry curvature sum rule is the key factor
determining the low frequency limit of the Hall conductance, ${\rm Re}[\sigma_{xy}(0)]$,
while the $f$-sum rule is the key factor in determining the Kerr effect at
large, optical, frequencies.   Since the commutator $\langle [ \hat{j}_x,\hat{j}_y] \rangle$ is strongly dependent on the intra-orbital pairing interaction 
$\Delta_{ab}$ in our model, we can conclude that intra-orbital pairing 
is the essential factor required for the observation of a significant
Kerr effect at optical frequencies, a conclusion reached earlier by Mineev~\cite{mineev2012}.

\section{Summary and Discussion}\label{sec:summary}

In this paper we have reviewed recent progress in understanding the
superconducting state of Sr$_2$RuO$_4$, mainly concentrating on
the Cooper pair spin as revealed, by the Knight shift~\cite{pustogow2019}, and the TRSB 
specifically focussing on its measurement by the Kerr effect~\cite{xia2006}. 
We have presented new calculations of the Kerr effect for the case of a model
$E_g$ spin singlet pairing state, of $d_{xz}+id_{yz}$ type. For this pairing state
we find a Hall conductance spectrum, as shown in Fig.~\ref{fig:rys3}, which is
qualitatively similar to that calculated earlier under the assumption of a chiral $p$-wave pairing state~ \cite{gradhand2013}.  We have shown that the Hall spectrum can be conveniently characterized in terms of two different sum rules. 
The first of these sum rules relates the low frequency limit of the
real Hall conductance, ${\rm Re}[\sigma_{xy}(\omega)]$, to Berry curvatures
of the Bogoliubov quasiparticle states~\cite{gradhand2014}.  The second
$f$-sum rule relates the high frequency limit of the real Hall conductance
${\rm Re}[\sigma_{xy}(\omega)]$ to the thermal expectation value of the equal-time current-current 
commutator, $\langle [\hat{j}_x,\hat{j}_y] \rangle$.  This commutator
is directly related to the order parameter for inter-orbital pairing, $\Delta_{ab}$
in our model, where $a = d_{yz}$ and $b=d_{zx}$.  This implies that observation
of a significant Kerr angle at optical frequencies shows the importance of
inter-orbital pairing in the superconducting state.

Of course, many open questions remain unanswered.  Firstly, we have not yet shown whether
a similar Kerr effect will occur in other candidate singlet pairing states,
such as $d+is$, $d+id$ or $d+ig$. The $E_g$ pairing state we considered
is the only spin-singlet pairing state with a symmetry enforced degeneracy at $T_c$,
and therefore the only one where the temperature of TRSB must be exactly equal to
the critical temperature, $T_c$.  General 
symmetry analysis requires that all other possible spin singlet TRSB pairing states 
are only strictly possible if there are two distinct phase transitions, $T_c$
 and then a lower one where TRSB occurs, $T_{TRSB}$ ~\cite{annett1990}. However,
 despite this, it has been argued that states such as $d+ig$~\cite{kivelson2020} can result from a near ``accidental'' degeneracy of the two distinct order parameters
 associated with different irreducible representations of the tetragonal
point group $D_{4h}$.  Support for this possibility is provided by the uniaxial 
strain experiments of Grinenko {\it et al.}~\cite{Grinenko2021} where distinct
$T_c$ and $T_{TRSB}$ transitions are found which separate as a function
of applied $\langle 100 \rangle$ uniaxial strain. Within experimental
accuracy these transitions are at the same temperature in unstrained samples, 
but clearly become distinct as a function of strain.  Qualitatively similar
behaviour might be expected in the $E_g$ pairing state considered above, but in that
case the splitting $T_c-T_{TRSB}$ should be linear in strain, while the
observed splitting is found to be quadratic. 
 
In conclusion, the precise pairing state of Sr$_2$RuO$_4$ still remains
unknown, despite over three decades of intensive experimental and theoretical
study~\cite{maeno2001,mackenzie2003,bergemann2003,kallin2012,maeno2012,kallin2016,mackenzie2017,huang2021,leggett2021,maeno2024}.  In many respects this system should be simpler to understand than the much more complex classes of materials such as cuprates and nicklates etc.  Sr$_2$RuO$_4$ is unique among these wider classes of
oxide materials because of its stoichiometric crystal
structure, relatively free of defects, and its well understood Fermi liquid normal
state above $T_c$.  The fact that the pairing state remains elusive 
shows that even in such a well-understood material, there is difficulty
of uniquely predicting or experimentally determining states of
unconventional superconductivity.

\acknowledgements{We would like to thank Dr Martin Gradhand for discussions
and use of the computer code to calculate the Kerr effect. The work of KIW has been supported by the M. Curie-Sk\l{}odowska University and the National Science Center, Poland (``Weave'' programme) through grant no. 2022/04/Y/ST3/00061.}


\begin{thebibliography}{99}


\bibitem{muller1986} K. A. M\"uller, J. G. Bednorz,
 Z. Phys.B {\textbf{64}}, 189 (1986).
DOI: https://doi.org/10.1007/BF01303701

\bibitem{maeno1994}
Y. Maeno, H. Hashimoto, K. Yoshida, S. Nishizaki, T. Fujita, J. G. Bednorz, F. Lichtenberg,
Nature \textbf{372}, 532  (1994).
DOI: https://doi.org/10.1038/372532a0

\bibitem{cava1994}
R. J. Cava, B. Batlogg, K. Kiyono, H. Takagi, J. J. Krajewski, W. F. Peck, Jr., L. W. Rupp, Jr.,  C. H. Chen,
Phys. Rev. B \textbf{49}, 11890 (1994).
DOI: https://doi.org/10.1103/PhysRevB.49.11890

\bibitem{mackenzie2003}
A. P. Mackenzie,  Y.  Maeno,
Rev. Mod. Phys. \textbf{75}, 657 (2003).
DOI: https://doi.org/10.1103/RevModPhys.75.657


\bibitem{rice1995}
T. M. Rice and M. Sigrist, 
J. Phys.: Condens. Matter \textbf{7}, L643 (1995).
DOI: 10.1088/0953-8984/7/47/002 


\bibitem{Luke1998} Luke, G., Fudamoto, Y., Kojima, K. et al. Time-reversal symmetry-breaking superconductivity in Sr2RuO4. Nature 394, 558–561 (1998). https://doi.org/10.1038/29038

\bibitem{ishida1998}  K. Ishida, H. Mukuda, Y. Kitaoka, K. Asayama, Z. Q. Mao, Y. Mori,  Y. Maeno, 
{\it Nature} \textbf{396}, 658 (1998).
DOI: https://doi.org/10.1038/25315

\bibitem{xia2006}
Jing Xia, Yoshiteru Maeno, Peter T. Beyersdorf, M. M. Fejer,  Aharon Kapitulnik,
Phys. Rev. Lett. \textbf{97}, 167002 (2006).
DOI: https://doi.org/10.1103/PhysRevLett.97.167002 


\bibitem{pustogow2019}
A. Pustogow, Yongkang Luo, A. Chronister, Y.-S. Su, D. A. Sokolov, F. Jerzembeck, A. P. Mackenzie, C. W. Hicks, N. Kikugawa, S. Raghu, E. D. Bauer, S. E. Brown,
Nature, \textbf{574}, 72 (2019).
DOI: https://doi.org/10.1038/s41586-019-1596-2


\bibitem{chronister2021}
Aaron Chronister, Andrej Pustogow Naoki Kikugawa, Dmitry A. Sokolov, Fabian Jerzembeck, Clifford W. Hicks, Andrew P. Mackenzie, Eric D. Bauer,  Stuart E. Brown,
PNAS  \textbf{118}, e2025313118  (2021).
DOI: https://doi.org/10.1073/pnas.2025313118


\bibitem{ishida2020}
Kenji Ishida, Masahiro Manago, Katsuki Kinjo, Yoshiteru Maeno,
J. Phys. Soc. Jpn. \textbf{89}, 034712 (2020). 
DOI: https://doi.org/10.7566/JPSJ.89.034712


\bibitem{Petsch2020}
Petsch, et al., Phys. Rev. Lett. {\textbf 125}, 217004
(2020).
DOI: https://doi.org/10.1103/PhysRevLett.125.217004 

 
\bibitem{spalek2022}
J. Spa\l{}ek, M. Fidrysiak, M. Zegrodnik, A. Biborski,
Physics Reports \textbf{959}, 1 (2022).
DOI: https://doi.org/10.1016/j.physrep.2022.02.003



\bibitem{gingras2022}
O. Gingras, N. Allaglo, R. Nourafkan, M. Côté,  A.-M. S. Tremblay,
Phys. Rev. B \textbf{106}, 064513 (2022).
DOI: 10.1103/PhysRevB.106.064513



\bibitem{bussman2022}
Annette Bussmann-Holder, Hugo Keller, 
Condens. Matter \textbf{7}, 10 (2022). 
DOI: https://doi.org/10.3390/condmat7010010

\bibitem{Anderson1987} 
P. W. Anderson, 
Science {\textbf 235} 1196-1198 (1987)
DOI: 10.1126/science.235.4793.119

\bibitem{Maier2005} 
T. A. Maier, M. Jarrell, T. C. Schulthess, P. R. C. Kent, and J. B. White, Phys. Rev. Lett. {\textbf 95}, 237001 (2005).
DOI: https://doi.org/10.1103/PhysRevLett.95.237001

\bibitem{Deng2016}
Xiaoyu Deng, Kristjan Haule, and Gabriel Kotliar,
Phys. Rev. Lett. {\textbf 116}, 256401 (2016)
DOI: https://doi.org/10.1103/PhysRevLett.116.256401

\bibitem{Kaser2022}
Stefan Käser, Hugo U. R. Strand, Nils Wentzell, Antoine Georges, Olivier Parcollet and Philipp Hansmann
Phys. Rev. B {\textbf 105}, 155101 (2022).
DOI: https://doi.org/10.1103/PhysRevB.105.155101

 
\bibitem{Romer2022}
Astrid T. Rømer, T. A. Maier, Andreas Kreisel, P. J. Hirschfeld and Brian M. Andersen,
Phys. Rev. Research {\textbf 4}, 033011 (2022)
DOI: https://doi.org/10.1103/PhysRevResearch.4.033011

\bibitem{maeno2001}
Yoshiteru Maeno, T. Maurice Rice, Manfred Sigrist,
Physics Today \textbf{54}, 42 (2001).
DOI: https://doi.org/10.1063/1.1349611

\bibitem{maeno2024}
Yoshiteru Maeno, Shingo Yonezawa, Aline Ramires,
J. Phys. Soc. Jpn. \textbf{93}, 062001 (2024).\\ 
DOI: https://doi.org/10.7566/JPSJ.93.062001

\bibitem{bergemann2003}
C. Bergemann, A. P. Mackenzie, S. R. Julian, D. Forsythe, E. Ohmichi,
 Advances in Physics, \textbf{52}, 639 (2003). 
 DOI: https://doi.org/10.1080/00018730310001621737






\bibitem{kallin2012}
Catherine Kallin,
 Rep. Prog. Phys. \textbf{75}, 042501 (2012).
DOI: 10.1088/0034-4885/75/4/042501


\bibitem{maeno2012}
Yoshiteru Maeno, Shunichiro Kittaka, Takuji Nomura, Shingo Yonezawa1,  Kenji Ishida,
J. Phys. Soc. Jpn. \textbf{81}, 011009 (2012).
DOI: https://doi.org/

\bibitem{kallin2016}
Catherine Kallin, John Berlinsky,
Rep. Prog. Phys. \textbf{79}, 054502 (2016).
DOI: 10.1088/003,4885/79/5/054502,


\bibitem{mackenzie2017}
Andrew P. Mackenzie, Thomas Scaffidi, Clifford W. Hicks, Yoshiteru Maeno, 
 npj Quant Mater \textbf{2}, 40 (2017). 
 DOI: https://doi.org/10.1038/s41535-017-0045-4

\bibitem{huang2021}
Wen Huang,
Chin. Phys. B, \textbf{30}, 107403 (2021).
DOI: 10.1088/1674-1056/ac2488

\bibitem{leggett2021}
Anthony J. Leggett, Ying Liu,
Journal of Superconductivity and Novel Magnetism  \textbf{34}, 1647 (2021).
DOI: https://doi.org/10.1007/s10948-020-05717-6

\bibitem{mackenzie1998}
A. P. Mackenzie, R. K. W. Haselwimmer, A. W. Tyler, G. G. Lonzarich, Y. Mori, S. Nishizaki,  Y. Maeno,
Phys. Rev. Lett. \textbf{80}, 161 (1998); [Erratum Phys. Rev. Lett. \textbf{80}, 3890 (1998)].
DOI: https://doi.org/10.1103/PhysRevLett.80.161



\bibitem{mackenzie1996}
A. P. Mackenzie, S. R. Julian, A. J. Diver, G. J. McMullan, M. P. Ray, G. G. Lonzarich, Y. Maeno, S. Nishizaki,  T. Fujita,
Phys. Rev. Lett. \textbf{76}, 3786 (1996).
DOI: https://doi.org/10.1103/PhysRevLett.76.3786



\bibitem{Curran2014}
P. J. Curran, S. J. Bending, W. M. Desoky, A. S. Gibbs, S. L. Lee, and A. P. Mackenzie,
Phys. Rev. B \textbf{89}, 144504 (2014)
DOI: https://doi.org/10.1103/PhysRevB.89.144504 

\bibitem{Curran2023} 
Curran, P.J., Bending, S.J., Gibbs, A.S. et al., Sci Rep 13, 12652 (2023). 
DOI: https://doi.org/10.1038/s41598-023-39590-9

\bibitem{kivelson2020}
Steven Allan Kivelson, Andrew Chang Yuan, Brad Ramshaw, Ronny Thomale,
npj Quantum Materials,  \textbf{5}, 43 (2020). 
DOI: https://doi.org/10.1038/s41535-020-0245-1


\bibitem{hasinger2017}
E. Hassinger, P. Bourgeois-Hope1, H. Taniguchi, S. René de Cotret, G. Grissonnanche, M. S. Anwar, Y. Maeno, N. Doiron-Leyraud1,  Louis Taillefer,
Phys. Rev. X \textbf{7}, 011032 (2017).
DOI: https://doi.org/10.1103/PhysRevX.7.011032



\bibitem{leggett1975}
Anthony J. Leggett
Rev. Mod. Phys. \textbf{47}, 331 (1975) [Erratum Rev. Mod. Phys. \textbf{48}, 357 (1976).]
DOI: https://doi.org/10.1103/RevModPhys.47.331 



\bibitem{murakawa2004}
H. Murakawa, K. Ishida, K. Kitagawa, Z. Q. Mao1, Y. Maeno,
Phys. Rev. Lett. {\bf 93} 167004 (2004).
DOI: https://doi.org/10.1103/PhysRevLett.93.167004 


\bibitem{annett1990} 
James F. Annett, 
 Adv. Phys. \textbf{39}, 83 (1990).
https://doi.org/10.1080/00018739000101481

\bibitem{annett2008}
James F. Annett, B. L. Györffy, G. Litak, K. I. Wysokiński,
Phys. Rev. B \textbf{78}, 054511 (2008). 
DOI: 10.1103/PhysRevB.78.054511

\bibitem{Duffy2000}
J. A. Duffy, S. M. Hayden, Y. Maeno, Z. Mao, J. Kulda4 and G. J. McIntyre
Phys. Rev. Lett. {\textbf 85}, 5412 (2000)
DOI: https://doi.org/10.1103/PhysRevLett.85.5412 



\bibitem{matsuki2026}
Hisakazu Matsuki, Rustem Khasanov, Jonas A. Krieger, Thomas J. Hicken, Kosuke Yuchi, Jake S. Bobowski, Giordano Mattoni, Atsutoshi Ikeda, Ryutaro Okuma, Hubertus Luetkens, Yoshiteru Maeno,
Phys. Rev. Lett. \textbf{136}, 066001 (2026).
DOI: 10.1103/sgcz-9rc7

\bibitem{Gupta2020}
R. Gupta, T. Saunderson, S. Shallcross, M. Gradhand, J. Quintanilla, and 
J. Annett,
Phys. Rev. B {\text bf 102}, 235203 (2020). 
DOI: https://doi.org/10.1103/PhysRevB.102.235203 


\bibitem{romer2019}
A. T. Romer, D. D. Scherer, I. M. Eremin, P. J. Hirschfeld, B. M. Andersen,
Phys. Rev. Lett. \textbf{123}, 247001 (2019).
DOI: 10.1103/PhysRevLett.123.247001


\bibitem{Gupta2022}
R. Gupta, S. Shallcross, J. Quintanilla, M. Gradhand and J. Annett,
Phys. Rev. B {\textbf 106}, 115126 (2022)
DOI: https://doi.org/10.1103/PhysRevB.106.115126 


\bibitem{ghosh2021}
Sudeep Kumar Ghosh et al 2021 J. Phys.: Condens. Matter \textbf{33} 033001.
DOI: https://iopscience.iop.org/article/10.1088/1361-648X/abaa06

\bibitem{wysokinski2019}
Karol Izydor Wysoki\'nski,
Condens. Matter \textbf{4}, 47 (2019). 
DOI:10.3390/condmat4020047



\bibitem{luke1998} 
G.Luke, Y.  Fudamoto,K. M. Kojima, M. I. Larkin, J. Merrin, B. Nachumi, Y. J. Uemura, Y. Maeno, Z. Q. Mao, Y. Mori, H. Nakamura,  M. Sigrist, {\it Time-reversal symmetry-breaking superconductivity in Sr$_2$RuO$_4$,}
Nature \textbf{394}, 558 (1998).
DOI: https://doi.org/10.1038/29038


\bibitem{capelle1997}
K. Capelle  E. K. U. Gross, B. L. Györffy,
Phys. Rev. Lett. \textbf{78}, 3753 (1997).
DOI: https://doi.org/10.1103/PhysRevLett.78.3753 

\bibitem{capelle1998}
K. Capelle  E. K. U. Gross and B. L. Györffy
Phys. Rev. B \textbf{58}, 473 (1998)
DOI: https://doi.org/10.1103/PhysRevB.58.473 

\bibitem{taylor2012}
Edward Taylor and Catherine Kallin
Phys. Rev. Lett. \textbf{108}, 157001 (2012).
DOI: https://doi.org/10.1103/PhysRevLett.108.157001

\bibitem{wysokinski2012}
K. I. Wysokiński, James F. Annett, and B. L. Györffy
Phys. Rev. Lett. \textbf{108}, 077004 (2012).
DOI: https://doi.org/10.1103/PhysRevLett.108.077004 



\bibitem{gradhand2013}
Martin Gradhand, Karol I. Wysokinski, James F. Annett, and Balazs L. Györffy
Phys. Rev. B \textbf{88}, 094504 (2013)
DOI: https://doi.org/10.1103/PhysRevB.88.094504 



\bibitem{mineev2007}
Vladimir P. Mineev
Phys. Rev. B \textbf{76}, 212501 (2007).
DOI: https://doi.org/10.1103/PhysRevB.76.212501 


\bibitem{mineev2012}
Vladimir P. Mineev
J. Phys. Soc. Jpn. \textbf{81}, 093703 (2012). 
DOI: https://doi.org/10.1143/JPSJ.81.093703

\bibitem{mineev2014}
V.P. Mineev
Phys. Rev. B \textbf{89}, 134519  (2014).
DOI: https://doi.org/10.1103/PhysRevB.89.134519 

 \bibitem{gradhand2014}
Martin Gradhand and James F Annett,  
J. Phys.: Condens. Matter \textbf{26}, 274205 (2014).
DOI: 10.1088/0953-8984/26/27/274205

\bibitem{goryo2007}
Jun Goryo
Phys. Rev. B \textbf{78}, 060501(R) (2008).
DOI: https://doi.org/10.1103/PhysRevB.78.060501 

\bibitem{lutchyn2009}
Roman M. Lutchyn, Pavel Nagornykh, and Victor M. Yakovenko
Phys. Rev. B \textbf{80}, 104508 (2009).
DOI: https://doi.org/10.1103/PhysRevB.80.104508 


\bibitem{konig2017}
Elio J. König and Alex Levchenko
Phys. Rev. Lett. \textbf{118}, 027001 (2017)
DOI: https://doi.org/10.1103/PhysRevLett.118.027001 

\bibitem{Suh2020}
H. G. Suh, H. Menke, P. M. R. Brydon, C. Timm, A. Ramires and 
D. F. Agterberg,
Phys. Rev. Research {\textbf 2}, 032023(R) (2020).
DOI: 10.1103/PhysRevResearch.2.032023

\bibitem{Lange1999}
E. Lange and G. Kotliar, 
Phys. Rev. Lett. {\text bf 82}, 1317 (1999).
DOI: https://doi.org/10.1103/PhysRevLett.82.1317 

\bibitem{Zhang2026}
Y. Zhang and H. Huang,
 arXiv:2604.08043
DOI: https://doi.org/10.48550/arXiv.2604.08043

\bibitem{Grinenko2021}
Grinenko, V., Ghosh, S., Sarkar, R. et al. 
Nat. Phys. {\textbf 17}, 748–754 (2021). DOI: https://doi.org/10.1038/s41567-021-01182-7

\end{thebibliography}

\end{document}